\documentclass[twocolumn,aps,prl,floatfix,showpacs]{revtex4}

\usepackage{amsmath}
\usepackage[dvips]{graphicx}
\usepackage{hhline}
\usepackage{epsfig}

\newcommand{\beq}{\begin{equation}}
\newcommand{\eeq}{\end{equation}}
\newcommand{\beqn}{\begin{eqnarray}}
\newcommand{\eeqn}{\end{eqnarray}}

\newcommand{\onbb}{$0\nu\beta\beta\,$}

\graphicspath{{pict/}}

\begin{document}

\title{Accurate shell-model nuclear matrix elements for neutrinoless 
double-$\beta$ decay}

\author{R.A.~Sen'kov and M.~Horoi}

\affiliation{Department of Physics, Central Michigan University, 
Mount Pleasant, Michigan 48859, USA}


\begin{abstract}
We investigate a novel method of accurate calculation of the 
neutrinoless double-$\beta$ decay shell-model nuclear matrix elements 
for the experimentally relevant case of $^{76}$Ge.  
We demonstrate that with the new method the nuclear matrix elements have 
perfect convergence properties and, using only the first 100 intermediate 
states of each spin, the matrix elements can be calculated with better 
than 1\% accuracy.
Based on {the analysis of neutrinoless double-$\beta$ decays of} $^{48}$Ca, 
$^{82}$Se, and $^{76}$Ge isotopes, we propose a new method to estimate the 
optimal values of the average closure energies at which the closure 
approximation gives the most accurate nuclear matrix elements. 
We also analyze the nuclear matrix elements for the heavy-neutrino-exchange 
mechanism, and we show that our method can be used to quench contributions 
from different intermediate spin states.
\end{abstract}

\maketitle


Observation of neutrinoless double-$\beta$ (\onbb) decay will have 
profound implications in modern physics. It will prove that 
the 
neutrino and antineutrino are identical particles (Majorana fermions), provide 
evidence 
of 
lepton-number violation, and help to determine the 
absolute scale of neutrino masses. In other words, it will change our 
understanding of Nature significantly.

In this 
paper,
we analyze the \onbb decay of $^{76}$Ge 
in a shell-model approach. From an experimental point of view, 
$^{76}$Ge is one of the most promising and important \onbb decay 
candidates. The most sensitive limits on \onbb decay half-lives have 
been obtained from germanium-based experiments: the Heidelberg-Moscow 
experiment~\citep{hm01}, the International Germanium 
experiment~\cite{igex02}, and the GERDA-I experiment~\citep{gerda13}. 
$^{76}$Ge is the only isotope for which an observational claim has been 
made (though it was not accepted by the double-beta decay 
community)~\citep{hm04,hm06}. GERDA-II~\cite{gerda04} and MAJORANA 
DEMONSTRATOR \cite{mdem13}, the second generation of the germanium-based 
experiments, are in progress.

Interpretation of the experimental results and planning of new experiments
require an accurate analysis of the \onbb decay process and the 
corresponding nuclear matrix elements (NMEs). Various theoretical models
have been used for NME calculations, including the quasiparticle 
random phase approximation (QRPA)~\cite{ves12,faess11,mika13}, the
interacting shell model (ISM)~\cite{prl100,prc13}, the interacting boson 
model (IBM-2)~\cite{iba-2}, the generator coordinate method 
~\cite{gcm}, and the projected hartree-fock bogoliubov model 
~\cite{phfb}. 

A \onbb decay process can be presented as a transition from the ground 
state of an initial nucleus to an arbitrary state of the intermediate 
nucleus and then transition to the ground state of the final nucleus. 
In an exact approach one needs to calculate all the intermediate states 
which can be a very demanding task. 
To avoid this computational challenge the closure approximation is 
usually introduced \cite{closure}. In the closure approximation the 
energies of the intermediate states are replaced with a constant value 
(closure energy), which allows the use of completeness to remove the 
sum over the intermediate states so that no information about the 
intermediate states is required.

In this 
paper, we present calculations of the NMEs for 
\onbb decay of $^{76}$Ge in the shell-model approach beyond closure 
approximation.
Going beyond closure requires the knowledge of a large number of the 
intermediate nuclear states. In the case of $^{76}$Ge, the intermediate 
nucleus $^{76}$As, being considered in a realistic model space, has 
about $1.5\times10^8$ states. Shell-model calculations for such a large 
number of nuclear states are practically impossible. 
%
%
To avoid the unmanageable computational costs, we use the mixed 
method~\citep{sh13,sh14}, in which the intermediate states are ordered 
according to their energies and a state cutoff parameter $N$ is 
introduced, so that all the intermediate states below the cutoff parameter
are taken into account exactly, i.e., in a nonclosure manner. The states 
which are above the cutoff
are included within the closure approximation. Defined in such a way, the 
mixed NMEs depend on both the state cutoff parameter $N$ and the closure 
energy. The mixed method was carefully tested in the fictitious cases 
of $^{44}$Ca and $^{46}$Ca, where all the intermediate states can be 
obtained, and then in the realistic case of $^{48}$Ca, where it is 
possible to get the first 500 intermediate states for each spin and 
parity $J^\pi$~\cite{sh13}. It was shown that the mixed NMEs converge 
much much more rapidly with an increasing state cutoff parameter $N$ compared 
to the nonclosure matrix elements. It was also shown that the mixed 
NMEs have very weak dependence on the closure energy, which makes this 
method more accurate compared to the closure approximation. Finally, 
the mixed method was successfully used to calculate the \onbb decay 
of $^{82}$Se where the first 250 intermediate states for each $J^\pi$ 
were calculated~\cite{sh14}.  
It was shown that in order to achieve a 1\% accuracy in the \onbb NME 
it is possible to consider only a small number of intermediate 
states: one needs about 20 intermediate states for each $J^\pi$ for the 
\onbb decay of $^{48}$Ca  and about 60 states for the \onbb decay of 
$^{82}$Se, while the corresponding total numbers of intermediate 
states for these cases are about $10^5$ and $10^7$. For $^{76}$Ge, 
about 100 intermediate states of nucleus $^{76}$As for each $J^\pi$ are 
required to provide a 1\% accuracy for the \onbb matrix elements. 
In calculations we use the shell-model code NuShellX@MSU~\cite{nushellxmsu}; 
the  $jj44$ model space, which consists of the nucleus $^{56}$Ni as a core 
and the $f_{5/2},\, p_{3/2},\, p_{1/2}$, and $g_{9/2}$ single-particles 
orbitals; and the JUN45 effective interaction \cite{jun45}. 

We demonstrate that the mixed method allows us to obtain practically exact  
values for the \onbb NMEs in the sense of going beyond the closure 
approximation. There are still uncertainties associated, for example, 
with the way the shell model treats the short-range correlations (SRCs), 
the restriction of the model space, and the effective interaction. 
However, since we know the exact (beyond closure) NMEs we can compare 
them with the closure NMEs and find optimal values for the average 
closure energies at which the closure approximation provides the most 
accurate NMEs. We have also calculated the optimal closure energies 
for the \onbb decays of $^{48}$Ca, $^{82}$Se, and $^{76}$Ge isotopes. 
One can expect a 7-10\% growth in the absolute values of the closure 
NME using our optimal closure energies instead of the commonly accepted 
ones~\cite{tomoda}.
We also discuss the contributions of the heavy-neutrino-exchange mechanism 
to the \onbb decay rate of $^{76}$Ge~\cite{ves12,prc13,prl13}.

Assuming the light-neutrino-exchange mechanism, the decay rate of a 
\onbb decay process can be written as \cite{ves12}
\beq \label{nme0}
\left[ T^{0\nu}_{1/2}\right]^{-1} = G^{0\nu} | M^{0\nu} |^2 
\left(\frac{\langle m_{\beta \beta}\rangle}{m_e}\right )^2,
\eeq
where $ G^{0\nu} $ is the phase-space factor \cite{kipf12}, 
$ M^{0\nu} $ is the NME, $m_e$ is the electron mass, 
and $\langle m_{\beta \beta}\rangle$ is the effective neutrino mass, 
which depends on the neutrino masses $m_k$ and the elements of the neutrino
 mixing matrix $U_{ek}$~\cite{ves12}.
The NME $M^{0\nu}$ is usually presented 
as a sum of three terms: Gamow-Teller ($M^{0\nu}_{GT}$), Fermi 
($M^{0\nu}_{F}$), and tensor ($M^{0\nu}_{T}$) NMEs 
(see, for example, Refs.~\cite{sh13}, \cite{sh14}, and \cite{prc10}). 

In the case of \onbb decay of $^{76}$Ge, the matrix elements 
can be presented as an amplitude for the transitional process where 
the ground state $|i\rangle$ of the initial nucleus $^{76}$Ge changes 
into an intermediate state $|\kappa\rangle$ of the nucleus $^{76}$As 
and then to the ground state $|f\rangle$ of the final nucleus $^{76}$Se,     
\beq \label{nme2}
M^{0\nu}_{\alpha}=\sum_{\kappa} \sum_{1234} \langle 1 3 | 
{\cal O}_{\alpha} | 2 4\rangle
\langle f |  \hat{c}^\dagger_{3} {\hat{c}}_4 | \kappa \rangle
\langle \kappa |  \hat{c}^\dagger_{1} {\hat{c}}_2 | i \rangle.
\eeq
Here the sum over $\kappa$ spans all the intermediate states 
$|\kappa \rangle$, indices 1-4 correspond to the single-particle 
quantum numbers, the label $\alpha$ describes different terms in the total NME 
(\ref{nme0}): Gamow-Teller ($\alpha=GT$), Fermi ($\alpha=F$), and tensor 
($\alpha=T$). The operators ${\cal O}_{\alpha}$ carry all the details
of a \onbb decay process, and they explicitly depend on the 
intermediate-state energy $E_\kappa$, 
${\cal O}_{\alpha}={\cal O}_{\alpha}(E_0+E_\kappa)$,
through the energy denominators in perturbation theory. 
The actual form of the ${\cal O}_{\alpha}$ operators can be found in 
Ref.~\cite{sh13}. Here, we would like only to emphasize the energy 
dependence of these operators. The constant 
$E_0=\left[E_{gs}({}^{76}\mbox{As})-E_{gs}({}^{76}\mbox{Ge})\right] + 
Q_{\beta\beta}/2 \approx 1.943 \mbox{ MeV}$. 

Exact calculation of the NMEs (\ref{nme2}) can be problematic due to 
the sum over a large number of intermediate states. One way to proceed 
in this situation 
%
is to use the {\it closure} approximation, in which the energies of 
intermediate states are replaced by a constant value so that
${\cal O}_\alpha (E_0+E_\kappa) \rightarrow \tilde{{\cal O}}_\alpha 
\equiv {\cal O}_\alpha(\langle E \rangle)$,
where $\langle E \rangle$ is the closure energy. Values of 
$\langle E \rangle$ from Ref.~\cite{tomoda} are frequently used.
%

%
To go beyond the closure approximation, a {\it nonclosure} approach can 
be considered. In this approach, the sum over intermediate states $\kappa$ 
in Eq.~(\ref{nme2}) is restricted by a finite cutoff parameter $N$. 
The success of the nonclosure approach is defined by the convergence properties 
of NMEs as a function of $N$.
%
The nonclosure approach cannot be directly used for the heavier cases, 
such as \onbb decay of $^{82}$Se and $^{76}$Ge, where only a few hundred 
intermediate states of each spin $J$ can be calculated. 
%

In the {\it mixed} method, the intermediate states below the cutoff 
parameter $N$ are taken into account within the nonclosure approach, 
while the states above $N$ are included in the closure approach. 
For more details see Refs.~\cite{sh13} and \cite{sh14}.

%

   
The nonclosure approach allows us to calculate the \onbb decay NMEs for a 
fixed spin and parity $J^\pi$ of the intermediate states $|\kappa\rangle$,
\beq \label{jdec}
M^{0\nu}_{\alpha}(J)=\sum_{\kappa, \; J_\kappa = J} 
\langle 1 3 | {\cal O}_{\alpha} | 2 4\rangle
\langle f |  \hat{c}^\dagger_{3} {\hat{c}}_4 | \kappa \rangle
\langle \kappa |  \hat{c}^\dagger_{1} {\hat{c}}_2 | i \rangle,
\eeq 
where the sum over $\kappa$ spans all the intermediate states with a given 
spin and parity $J^\pi$. This $J$ decomposition can be obtained only 
within a nonclosure approach. 

We also analyze the NMEs for the right-handed 
heavy-neutrino-exchange mechanism, whose corresponding contribution to the total decay rate can be written as
\beq\label{hnme0}
\left[ T^{0\nu}_{1/2} \right]_{\mbox{heavy}}^{-1} = G^{0\nu} | M^{0\nu}_N |^2
|\eta_{NR}|^2,
\eeq
where the heavy-neutrino-exchange matrix elements $M^{0\nu}_N$ have a 
structure similar to that of the light-neutrino-exchange NMEs, while the parameter $\eta_{NR}$ depends on the heavy-neutrino masses (for more details see, for example, Ref. \cite{prc13}). One difference between the heavy- and the light-neutrino-exchange mechanisms is that the heavy-neutrino-exchange NMEs do not depend on the energy of intermediate states. Thus for the heavy-neutrino-exchange mechanism the closure approach provides the exact matrix elements. 

\begin{figure}
\begin{center}
\includegraphics[width=0.46\textwidth]{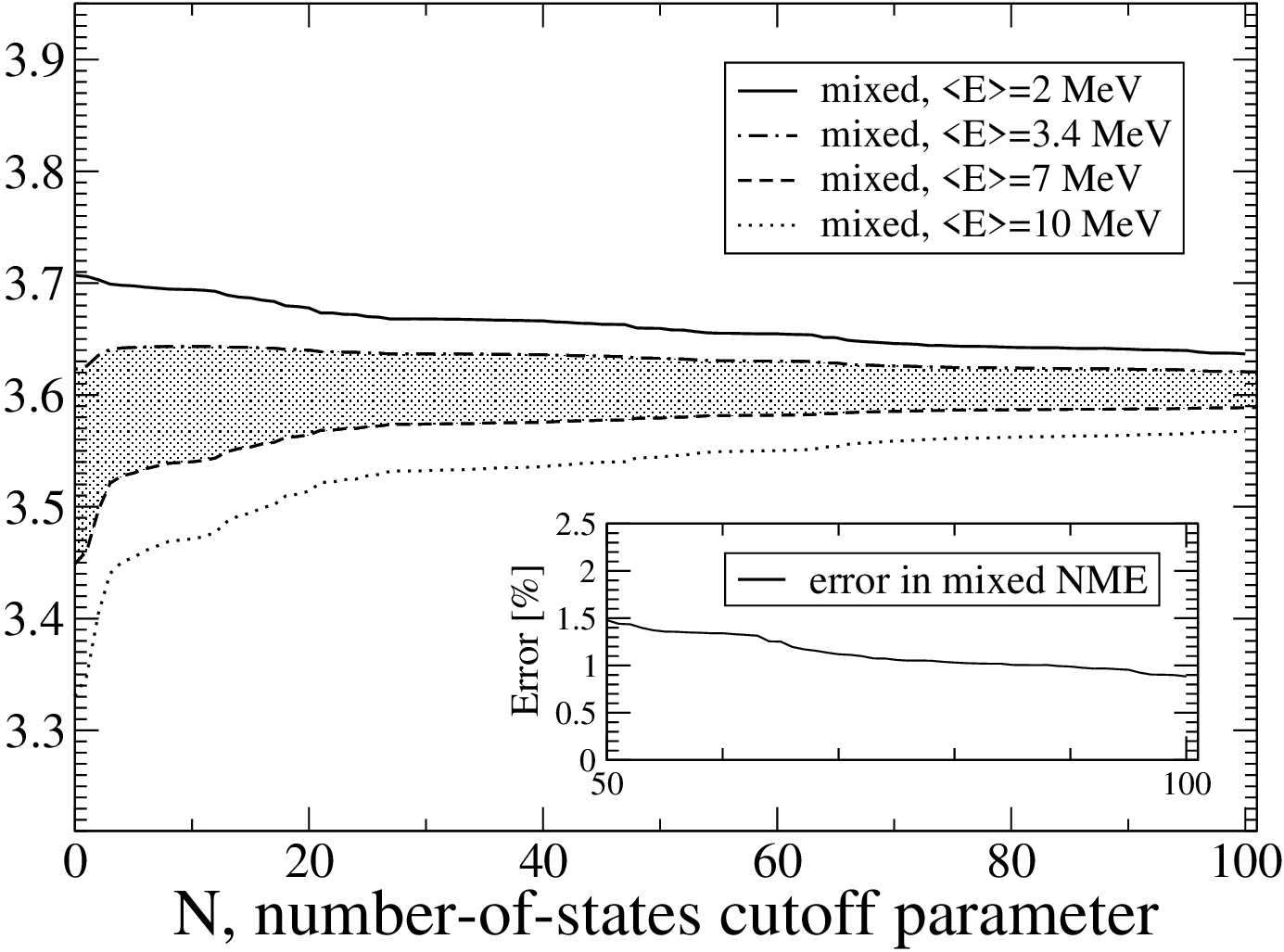}
\caption{Dependence of mixed NMEs (light-neutrino exchange) 
on the cutoff parameter $N$ calculated for different average closure 
energies $\langle E \rangle$. $\langle E \rangle=2$ MeV 
(solid curve), $\langle E \rangle=3.4$ MeV  (dash-dotted curve), 
$\langle E \rangle=7$ MeV (dashed curve), and $\langle E \rangle=10$ MeV 
(dotted curve). Inset: Uncertainty in the value of mixed 
NMEs corresponding to the shaded area.\\ \label{fig2}}
\end{center}
\end{figure}

\begin{figure}
\begin{center}
\includegraphics[width=0.46\textwidth]{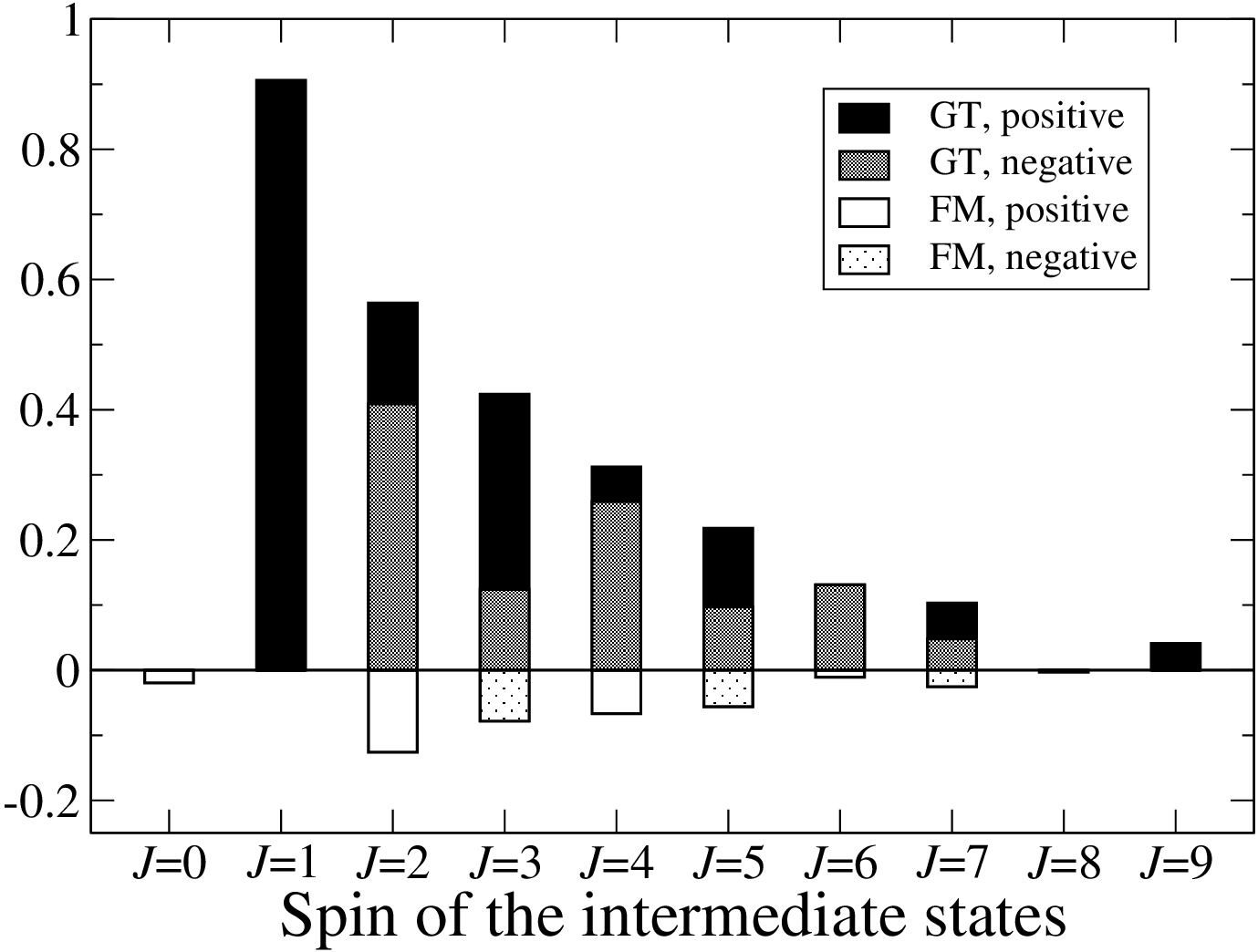}
\caption{$J$ decomposition: contributions of the intermediate states 
$|\kappa \rangle$ with a certain spin and parity $J^\pi$ to the nonclosure 
Gamow-Teller (dark colors) and Fermi (light colors) matrix elements for the 
$0\nu\beta\beta$ decay of ${}^{76}$Ge (light-neutrino exchange). Solid black 
and white bars correspond to positive-parity states, while shaded bars 
represent states with a negative parity. The CD-Bonn SRC parametrization was 
used. \label{fig4}}
\end{center}
\end{figure}


First, we studied the convergence properties of the \onbb decay NMEs 
of $^{76}$Ge. 
$N\!=\!100$ is the maximum number of states we are able to calculate in $^{76}$As with an computational effort of about 500000 $CPU \times hour$.
%
%
%
In the mixed method, the states above the cutoff parameter $N$ are included 
in the closure approximation, which makes the mixed NMEs dependent on 
the closure energy $\langle E \rangle$. However this dependence is not 
strong. For $N\!=\!0$ (the closure approximation), it results in a 10\% 
uncertainty in the total NMEs~\cite{prc10}. When the cutoff parameter 
increases, this dependence weakens relatively 
rapidly.  
Figure~\ref{fig2} shows the convergence properties of the mixed NMEs in an 
enhanced form and how these properties change when the closure energy varies. 
The solid, dash-dotted, dashed, and dotted lines in the figure present 
the mixed NMEs calculated with $\langle E\rangle$ equal to 2, 3.4, 7, 
and 10 MeV, respectively. If we restrict the range of possible closure 
energies to 3.4 to 7.0 MeV (which is quite reasonable since 
one curve approaches the final NME from above and the other approaches 
it from  below, so the true NMEs should be confined somewhere in between), 
then the corresponding shaded area 
gives us the 
uncertainty in the mixed NMEs. We can see how the uncertainty goes down 
when the cutoff parameter $N$ increases.  
The corresponding relative error in the mixed matrix elements is presented 
in the inset in Fig.~\ref{fig2}. It shows that it is sufficient to 
use only the first 100 nuclear states for each $J^\pi$ of $^{76}$As 
to obtain the \onbb decay NMEs of $^{76}$Ge within a 1\% accuracy.   

Figure~\ref{fig4} 
presents the $J$ decomposition [see Eq.~(\ref{jdec})] 
of the nonclosure NMEs. 
%
All the Gamow-Teller matrix elements are positive and all the Fermi matrix 
elements are negative. If we neglect the tensor NMEs (which are actually 
small), then the total height of each bar corresponds to the total NMEs  calculated for each spin $J$ in Eq.~(\ref{jdec}). 
We can see that all the spins contribute coherently to the total NMEs. 
The contribution of $J=1$ is dominating, but it provides only about 30\% 
of the total value. If we include only the $J=1$ intermediate states,  
then we will lose about 70\% of the total matrix elements and about 91\% 
of the decay rate.
%
%
Table~\ref{tbl2} summarizes the results for the light-neutrino-exchange NME 
 \onbb decay of $^{76}$Ge calculated for different SRC parametrization 
sets~\cite{prc10}. The mixed total matrix element is about 7\% percent 
greater than the total closure NME. This increase is consistent with  
similar calculations~\cite{sh13,sh14,qrpa-cl}. 


\begin{table}[ht]
\caption{Mixed and closure (last column) NMEs for the $0\nu\beta\beta$ 
decay of ${}^{76}$Ge (light-neutrino exchange) calculated with different 
SRC parametrization schemes~\cite{prc10}. Closure NMEs were calculated 
for a standard closure energy of $\langle E \rangle=9.41$ MeV \cite{tomoda}.
\label{tbl2}}
\begin{ruledtabular}
\begin{tabular}{lccccc}
 \; SRC \; & \; ${ M}^{0\nu}_{GT}$ \; &
\; ${ M}^{0\nu}_F$ \; & \; ${ M}^{0\nu}_T$ \; & 
${ M}^{0\nu}_{total}$ & ${\cal M}^{0\nu}_{closure}$ \\
\hline
 None & \; 3.06 \; & \; -0.63 \; & \; -0.01 \; & \; 3.45 \; &  3.24 \\
 Miller-Spencer &  2.45 & -0.44 & -0.01 & 2.72 & 2.55\\
 CD-Bonn & 3.15 & -0.67 & -0.01 & 3.57 & 3.35 \\
 AV18 & 2.98 & -0.62 & -0.01 & 3.37 & 3.15 \\
\end{tabular}
\end{ruledtabular}
\end{table}

It should be noted that the $jj44$ model space is incomplete because the  
$f_{7/2}$ and $g_{7/2}$ orbitals are missing. As a result the Ikeda sum 
rule is not satisfied and some contributions from the Gamow-Teller NME 
with $J^\pi=6^+$ and $8^+$ and from the Fermi NME $J^\pi\!=\!1^-$ are 
missing. Looking at Fig.~\ref{fig4}, it seems safe to suggest that the 
missing contributions are not very large. However, this deficiency is 
reflected in the two-neutrino NME, which requires a quenching factor 
of about 0.64, smaller than the usual 0.74, to describe the experimental 
data (see also Table 2 in Ref.~\cite{caurier12}). Although the spin-isospin 
operators entering the $0\nu\beta\beta$ decay NME are different from  
those in the 
pure Gamow-Teller, some authors (see, e.g., Ref.~\cite{ejiri13}) advocate 
using appropriate quenching factors for contributions coming from different
spins of the intermediate states. The most important are those from 
$J^{\pi}=1^+$ states, which represent about 30\% of the total NMEs, 
and from $J^{\pi}=2^-$ states \cite{ejiri13}, which represent 
about 15\% of the total NMEs. It would be interesting to investigate whether 
quenching factors obtained from other processes, such as $2\nu\beta\beta$ 
decay and charge-exchange reactions, quench the corresponding 
contributions to the $0\nu\beta\beta$ decay NMEs. For example, if one 
uses a quenching factor of $0.64^2$ for the contribution from the 
$J^{\pi}\!=\!1^+$ states and $0.40^2$ for the contribution from the 
$J^{\pi}=2^-$~\cite{ejiri13}, one gets for the CD-Bonn SRC an NME of 2.369 
rather than 3.572 (see Table I). One can view this as a lower limit 
NME in our approach.

\begin{figure}
\begin{center}
\includegraphics[width=0.46\textwidth]{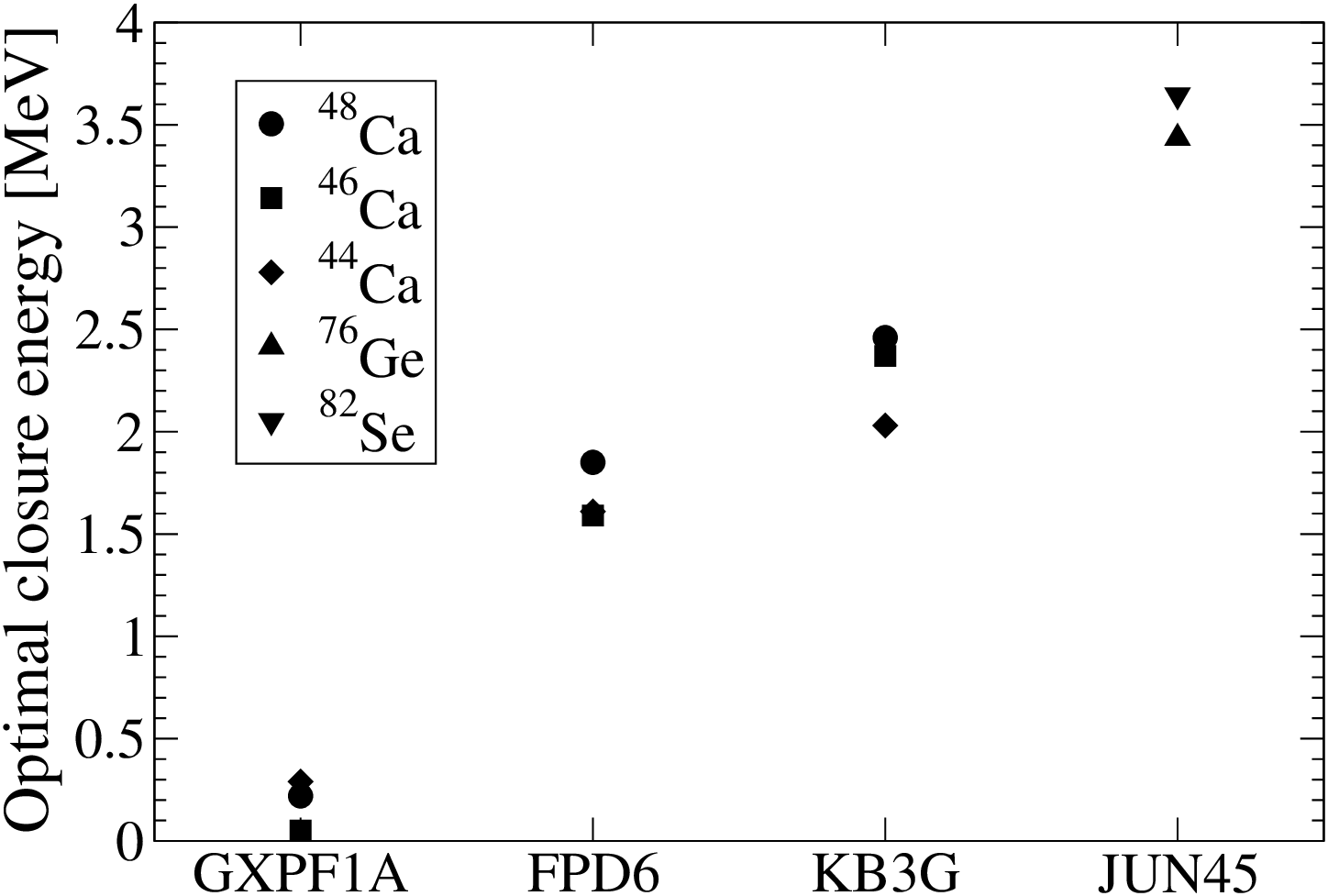}
\caption{Optimal closure energies $\langle E \rangle$ calculated for 
different isotopes and effective interactions. Fictitious \onbb decays:  
$^{44}$Ca (diamonds) and $^{46}$Ca (squares). Real decays: $^{48}$Ca 
(circles), $^{76}$Ge (upward triangles), and $^{82}$Se (downward triangles). 
Effective interactions considered are GXPF1A, FPD6, and  
KB3G for Ca and JUN45 for Ge and Se isotopes. \label{fig6}}
\end{center}
\end{figure}

Since we can calculate both the beyond-closure NME and the closure NME, 
it is possible to find such optimal values for the closure energies at 
which the closure approach provides the most accurate NMEs (see, e.g.,  
the crossing lines in Fig. 5 in Ref.~\cite{sh14}).
%
One interesting observation is that the optimal energies calculated for 
the \onbb decay of $^{82}$Se \cite{sh14} and $^{76}$Ge with the same 
JUN45 effective interaction and the same $jj44$ model space 
practically coincide: they both equal about $\langle E \rangle \approx 3.5$
MeV, although the two cases describe quite different nuclei. It would 
thus be interesting to find a method to estimate the optimal closure 
energies rather then using estimates from other methods, such as those 
in Ref.~\cite{tomoda}. 
Figure~\ref{fig6} presents the optimal closure energies calculated for 
the fictitious \onbb decays of $^{44}$Ca (diamonds) and $^{46}$Ca 
(squares) and for the realistic \onbb decays of $^{48}$Ca (circles), 
$^{76}$Ge (upward triangles), and $^{82}$Se (downward triangles). All calcium 
isotopes were calculated in the $pf$ model space using several 
realistic interactions. The $^{76}$Ge and $^{82}$Se isotopes were 
considered in the same $jj44$ model space and with the same JUN45 
interaction.
The optimal closure energies are significantly lower than the standard 
closure energies (7.72 MeV for Ca, 9.41 MeV for Ge, and 10.08 MeV for 
Se~\cite{tomoda}), which explains the 7--10\% growth in absolute values 
of the nonclosure NMEs compared to the closure values. We conjecture that 
the optimal energies depend on the effective interaction and, possibly,  
on the model space. We found the optimal closure energies for the three 
interactions in the $pf$ model space: GXPF1A~\cite{gxpf1a}, 
FPD6~\cite{fpd6}, and KB3G~\cite{kb3g}. 
However, it seems that the energies do not depend much on the specific 
nucleus: all the calcium isotopes calculated with the same interaction 
and both the $^{76}$Ge and the $^{82}$Se isotopes calculated with the same model 
space and with the same interaction give similar optimal closure energies.
This opens up an interesting opportunity: one could calculate the optimal 
closure energy in a realistic model space with an effective interaction 
for a nearby less computationally demanding isotope (for example, 
$^{44}$Ca), after which one could use it for a realistic case 
(for example, $^{48}$Ca). 
This scheme offers a consistent way of ``calculating" the closure 
energies that has not been discussed before. 

We also calculated the heavy-neutrino-exchange mechanism NMEs 
(see, e.g., Ref.~\cite{prc13} for more details) for the \onbb of $^{76}$Ge,
and we get a value of 202 for the CD-Bonn SRC and 126 for the AV18 SRC. 
Their $J^{\pi}$ decompositions will be published elsewhere \cite{shbig}.


Summarizing, we have calculated the \onbb decay NMEs of $^{76}$Ge using, for 
the first, time a realistic shell-model approach beyond closure 
approximation. 
%
%
We have demonstrated that the mixed NMEs converge very rapidly compared to the 
nonclosure matrix elements and we found a 7-10\% increase in the total 
NMEs compared to the closure values.

For the light-neutrino-exchange mechanism we predict 
$M^{0\nu}\!=\!3.5 \pm 0.1$ for \onbb decay of $^{76}$Ge, where the average 
value and the error were estimated considering the NMEs calculated with the  
CD-Bonn and AV18 SRC parametrization sets.  
These values should be compared with the corresponding calculations 
performed within different approaches:
2.96 (ISM-1~\cite{npa818}), 3.77 (ISM-2~\cite{ism-eff}), 4.6 (EDF~\cite{gcm}), 
2.28-4.17 (QRPA-Jy~\cite{suh2012}), and 5.42 (IBM-2~\cite{iba-2}).
For the heavy-neutrino-exchange NME for $^{76}$Ge, we get a value of 202 
for the CD-Bonn SRC and 126 for the AV18 SRC. The corresponding QRPA results are 412 and 265~\cite{ves12}, and the IBM-2 results are 163 and 107~\cite{iba-2}.

We have proposed a new method of calculating the optimal closure energies at 
which the closure approach gives the most accurate NMEs. We argue that 
these optimal closure energies depend on the interaction and model space 
and have a weak dependence on the actual isotopes. 
It offers the opportunity to estimate the beyond-closure \onbb NMEs without 
actually calculating the intermediate states.

We have calculated for the first time a decomposition of the shell-model NMEs 
in light- and heavy-neutrino-exchange mechanisms for different spins of 
intermediate states. We found that for the light-neutrino-exchange NMEs 
the contribution of the $J^{\pi}=1^+$ states is about 30\% and that 
of the $J^{\pi}=2^-$ states is about 15\%. 
The shell-model $J$ decomposition that we obtained provides a unique 
opportunity to selectively quench different contributions to the total 
NMEs, which, in the case of $^{76}$Ge, could lead to a decrease in the 
total matrix elements by about 30\%. 
Although the QRPA approach can provide a $J$ decomposition, its 
methodology of choosing the $g_{pp}$ parameter to describe the 
$2\nu\beta\beta$ half-life~\cite{qrpa-cl} makes the selective 
quenching ambiguous.


The authors thank B.A. Brown and V. Zelevinsky for useful discussions.
Support from  the NUCLEI SciDAC Collaboration under
U.S. Department of Energy Grant No. DE-SC0008529 is acknowledged. 
M.H. also acknowledges U.S. NSF Grant Nos. PHY-1068217 and PHY-1404442. 


\end{document}